\documentclass[onecolumn, preprint, aps,amsmath,amssymb,superscriptaddress]{revtex4}
\usepackage{dcolumn}% Align table columns on decimal point
\usepackage{bm}% bold math
\usepackage{amsmath}
\usepackage{amsfonts}
\usepackage{graphics}
\usepackage{graphicx}
\usepackage{enumerate}
\usepackage{color}   % {\color{red} TEXT }

\begin{document}

\title{Methods for measuring  the citations and productivity of scientists across time and discipline}
\author{Alexander M. Petersen}
\affiliation{Center for Polymer Studies and Department of Physics, Boston University, Boston, Massachusetts 02215, USA}
\author{Fengzhong Wang}
\affiliation{Center for Polymer Studies and Department of Physics, Boston University, Boston, Massachusetts 02215, USA}
%\author{Kevin Stokely}
%\affiliation{Center for Polymer Studies and Department of Physics, Boston University, Boston, Massachusetts 02215, USA}
\author{H. Eugene Stanley}
\affiliation{Center for Polymer Studies and Department of Physics, Boston University, Boston, Massachusetts 02215, USA}
\date{\today}

\begin{abstract} Publication statistics are  ubiquitous in the ratings of scientific achievement, with citation counts
and paper tallies factoring into an individual's  consideration for postdoctoral positions, junior faculty, and tenure. 
Citation statistics are designed to quantify individual career achievement, both at the level of a single publication,
and over an individual's  entire career. 
While some academic careers are defined by a few significant papers (possibly out of many), other academic careers are
defined by the cumulative contribution made  by the author's publications to the body of science. 
Several metrics have been formulated to quantify an individual's publication career, yet none
 of these metrics account for the collaboration group size, and the time dependence of citation counts. In this paper we
 normalize publication metrics in order to achieve a universal framework for analyzing
and comparing scientific achievement across both time and discipline. We study the  publication careers of individual authors over the 50-year period
1958-2008 within six high-impact journals:  {\it  CELL},  {\it the New England Journal of Medicine (NEJM)}, {\it
Nature}, {\it the 
Proceedings of the National Academy of Science (PNAS)}, {\it Physical Review Letters (PRL)},  and {\it Science}. 
 Using the normalized metrics (i) ``citation shares" to quantify scientific success, and (ii) ``paper
shares" to quantify scientific productivity, we compare the career achievement of individual authors within each
journal, where each journal represents a local arena for competition. 
We uncover quantifiable statistical regularity in the probability density function (pdf) of scientific achievement in
all journals analyzed, which suggests that a fundamental driving force underlying scientific achievement is the
competitive nature of scientific advancement. 
\end{abstract}
%\pacs{01.80.+b,89.75.Da,02.50.Fz}
\maketitle

\section{Introduction} 
 The study of human success is difficult because information has traditionally been recorded for only the excellent,
while individuals with lower than average careers are generally neglected in the record books. Hence, drawing
conclusions based only on the relatively few stellar careers will suffer to some extent from selection bias. 
In contrast,  conclusions drawn from the entire population might better illustrate the mechanisms of success.
 While it is not feasible 
to obtain the career  publication data for every scientist in every journal, it is possible to study a subset of
scientists that succeed in publishing in a specific journal.

Several empirical studies have analyzed the citation statistics of individual papers
\cite{howpopular, 110PhsRev, 100citedist,GooglePageRank, DiffusionRanking}, of individual journals/fields
\cite{DiffusionRanking, citationsPNAS, ScalingJournals, hindexFields, BrazilPhysChem, Jranking, IFactor}, and of subsets
of individuals
\cite{DiffusionRanking, SoretteStrechedExp, Hindex, hindexResearchers}. In this paper, we 
study the cumulative citation statistics of {\it individual} scientists over their publication careers within a given
journal. Studying the distribution of career accomplishment in a particular journal serves as a proxy for the more
difficult 
task of studying the citation statistics of all individuals in all journals, where such all-encompassing data are not as
readily available. Here we adopt the working hypothesis that studying publications in high-impact journals offers crude
approximation to an author's  scientific contribution. 

We  develop a simple method for normalizing
citations so that they can be compared across time and discipline. In order to compare across time, we normalize the number of citations for each paper in a given year
 by the average number of citations to papers from the same publication year. This
re-scaling can also aid in removing discipline-specific citation patterns that vary across discipline, especially when
considering discipline-specific journals.
We further  remove discipline-specific collaboration patterns by dividing the achievement equally among the collaboration
group members.

 This work aims to demonstrate the importance of properly normalizing  any conceivable 
metric that quantifies career achievement (e.g. the citation count, h-index). Extending the work of Radicchi {\it et
al.}
\cite{citationsPNAS}, which normalizes the citation values of single articles across discipline by re-scaling to local
citation averages,
our goal is  to  provide a framework for normalizing the scientific achievement of
individual careers. The methodology developed in this paper should conceivably make possible the comparison of
careers between various fields. Furthermore, we are able to
study
 the mechanisms of human success
in scientific arenas,  where effective competition  arises from limited financial, temporal, and creative resources.

In addition to studying the distribution of success and productivity, in this paper we also investigate the waiting
time between successive achievements, which is intrinsically related to the underlying mechanism of progress. 
Recent work in \cite{BB1, BB2}
demonstrates that the Matthew  effect (the ``rich--get--richer" effect) can be quantified by analyzing the career
longevity of employees within competitive professions, such as professional sports and academia. Here, we demonstrate 
the Matthew effect on the scale of individual authors by analyzing the time intervals between successive publications in
high-impact journals. The Matthew effect \cite{MatthewI} derives from a passage in  the Gospel of Matthew and is a
popular conceptual theory in sociology. 
This theory is analogous to several other  positive feedback or cumulative advantage theories
\cite{cumadvprocess, huberCum} which 
have been used to explain the ubiquity of right-skewed distributions that arise in socioeconomic studies. Of particular
note, the  generic preferential attachment mechanism  is relevant to the dynamics of citations
\cite{growingnetworks, ABPA} as well as the dynamics of human sexual networks \cite{sexPA}. \\

First, we briefly summarize several results which are relevant to the analysis of success and productivity performed in
this paper. A seminal study performed over 50 years ago by W.~ Shockley \cite{ShockleyProductivity} studied the rate of
productivity measured by the total number of publications and the total number of patents filed   at several large
research institutions. In this paper, Shockley   suggests that normalizing metrics for output (e.g. patents, papers,
citations) by the number of individuals could alleviate the discrepancy between disciplines. In this paper, we normalize
metrics for output by the number of contributors such that they are weighted ``shares", a procedure recently employed in
the calculation of h-index values
\cite{hindexResearchers}.

 Laherr\`ere {\it et al.} \cite{SoretteStrechedExp}  analyze the top 1120 most-cited physicists over the sixteen-year
period 1981-1997, with the result that the distribution of cumulative citation counts among these scientists, without
any normalization procedure, is described by a stretched exponential pdf. 

%After implementing our normalization procedure, 
%we observe heavy tails that are more consistent with an asymptotic power-law  pdf $P(x) \sim x^{-\alpha}$, with $\alpha
%\approx 2.5$ for 
%all journals analyzed.

Redner \cite{howpopular} analyzes approximately 800,000 individual papers and found that the pdf $P(x)$ of citations per
paper $x$ follows an approximate inverse-cubic power law. This result is 
found by analyzing the Zipf plot of the number of citations to a particular paper.  Interestingly, we find that
this result is  maintained even after the normalization procedure developed in this paper. Redner
\cite{110PhsRev} also analyzes 110 years of citation statistics in {\it Physical Review} journals, where he calculates
the citation distribution of $353,268$ papers,  and finds a log-normal pdf $P(x)$ (without normalizing for publication
time). 
In addition to the size and coverage of the  {\it Physical Review} database, another impressive feature of  this study
is the analysis of
 citation dynamics,  relating  the citation growth rate to the number of contemporaneous citations. Of particular note, Redner finds an
approximately linear citation (attachment) rate for citations originating from within {\it Physical Review}
publications.  Also, a recent study \cite{Jranking} analyzes the citation dynamics in
2,267 journals
and finds that the  time-dependent average number of citations per paper within each journal approaches a steady state value which can be
used as a normalizing factor for comparing journals across discipline. In this paper, we use the time-dependent average number of citations per
paper for a particular journal and publication year as the normalizing factor in order to compare articles across time
and discipline. 

Recently,  Hirsch \cite{Hindex} proposed the  h-index to be an unbiased  metric to quantify scientific impact. The
h-index is calculated using the raw number of citations for each of an author's papers.  Although simple in its
definition, the h-index has encountered scrutiny, with the opposition 
claiming that the definition of the h-index is biased in that it neglects differences in publication patterns between
scientific (sub)disciplines.
It is further biased in that it neglects variations in the size of collaborations, and hence, the credit associated
with a given publication.
In Ref.~\cite{hindexResearchers},
Batista {\it et al.} normalize the h-index to the number of authors contributing to each paper in order to 
account for differences in publication styles across discipline. Two additional  studies suggest that normalizing by
the size of the field
can alleviate the differences in research and
publication style across disciplines \cite{hindexFields, citationsPNAS},  however  the relationship between
citation trends and field size are not trivial and depend on several factors \cite{IFactor}.
Very recently, an enormous study of twenty-five
million papers by Wallace {\it et al.} \cite{100citedist} implements Tsallis statistics to investigate the distribution
of citations for papers spanning the 106-year period 1900-2006.  
The extensive analysis in \cite{100citedist} also  discusses the changes in citation trends  over time and the ``uncited''
phenomena. 

 Our main result is to provide the first study that quantifies the career publication statistics of individual authors
while normalizing the publication statistics with respect to 2 factors:
\begin{enumerate}[(i)] % Requires::  usepackage{enumerate}
\item the number of authors credited for a particular publication,  
\item the time-dependent increase of citation counts. 
\end{enumerate} Specifically, we account for (i) by normalizing citation counts and paper tallies to the number of
contributing authors, and (ii)   by normalizing citations by  the local average number of citations per
paper; the local average is computed from the set of papers published in the same journal in the same
year.
 While these two factors have been discussed, to our knowledge, no study has incorporated them both simultaneously. 

In order to compare careers that are of similar duration, we use methods described in
\cite{huber98} to isolate ``completed" careers from our journal data bases, which all span the 50-year period 1958-2008
except for CELL which was created only in 1974. For the subset of careers that meet a completion
criterion, we normalize each individual citation according to  factors (i) and (ii). 
We then tally the normalized citation shares for each scientist, which serves as one possible metric for career accomplishment.
We also perform the analogous 
procedure for paper shares which serves as a metric for career productivity.\\

The organization of this paper is as follows: in Section \ref{section:DataMethods} we review the data analyzed, the
procedure with which we aggregate the data into publication careers, and the possible systematic errors inherent in our
method. In Section \ref{section:Results} we analyze the distribution of both citation and productivity statistics for
three  high-impact multidisciplinary journals: {\it Nature},  {\it the 
Proceedings of the National Academy of Science (PNAS)}, and {\it Science}, and also for three less multidisciplinary
journals: {\it CELL},
 {\it the New England Journal of Medicine (NEJM)}, and { \it Physical Review Letters (PRL).}  We note that
  only three of these journals analyzed are discipline specific, and so we rely significantly on the results of
\cite{hindexFields, citationsPNAS} in justifying the comparison of normalized career metrics across discipline beyond
our results for the high-impact journals {\it CELL},
 {\it the New England Journal of Medicine (NEJM)}, and { \it Physical Review Letters (PRL). }
\section{Data and Methods}
\label{section:DataMethods}

We downloaded journal data  in May 2009 from ISI Web of Knowledge \cite{WofK}. We restrict our analysis to publications
termed ``Articles," which excludes reviews, letters to editors, corrigendum, etc. 
For each journal, we combine all publications into one database. In total, these data  represent approximately 350,000
articles and  600,000 scientists (see Table \ref{table:journals}).

\begin{table}
\caption{ Summary of data set size for each journal. Total number N of unique (but possibly degenerate) name
identifications.}
\begin{tabular}{@{\vrule height 10.5pt depth4pt  width0pt}lc||c||c||}\\
%&\multicolumn4c{Least-square values}&\multicolumn3c{gamma distribution moments}\\
\noalign{
\vskip-11pt}
\vrule depth 6pt width 0pt \textbf{\em Journal}  &  Years & Articles & Authors, N  \\
\hline  
CELL &  \  \ 1974-2008 \ \  & \ \ 53,290 \ \ &  \ \ 31,918 \ \  \\
NEJM &  1958-2008 & 17,088 &  66,834 \\
Nature &  1958-2008 & 65,709 &  130,596 \\
PNAS &  1958-2008 & 84,520 &  182,761 \\
PRL &  1958-2008 & 85,316 & 112,660 \\
Science &  1958-2008 & 48,169 & 109,519 \\
\hline
\end{tabular}
\label{table:journals}
\end{table}

Our data collection procedure begins with downloading all ``articles" for each journal for  year $y$ from ISI Web of
Knowledge. 
From the set of $N(y)$ articles for each particular journal and year, we calculate $\langle c(y)\rangle$, the average
 number of citations per article at the date of data extraction (May 2009). Each article summary includes a field for a
contributing author's name identification, which consists of a last name and first and middle initial \cite{authorname}.
From these fields, we aggregate the career works of individual authors within a particular journal. In this paper we
develop normalized  metrics for career success and productivity, while in \cite{BB2} we compare theory and empirical
data for career longevity.

For each author, we combine all his/her articles in a given journal.  Specifically, a publication career in this
paper 
refers to the lifetime achievements of a single author within a single journal, and not the lifetime achievements
combined among the six journals analyzed. 
 We define $n$ as the total number of papers for a given author
 in a given journal over the 50-year period.  In analogy with the traditional citation tally, 
 one can calculate the career success/impact within a given journal
 by adding together the citations $c_{i}$ received by the $n$ papers,
\begin{equation}
C=\sum_{i=1}^{n}c_{i} \ .
\label{totalcitations}
\end{equation}
Furthermore, one can calculate the 
career productivity of a given author within a specific journal as the total number $P$ of papers published within the
journal. A main point raised in this paper is to discount the value of citation metrics which do not take into account
the time-evolution of citation accumulation.

\begin{figure}
\centering{\includegraphics[width=0.9\textwidth]{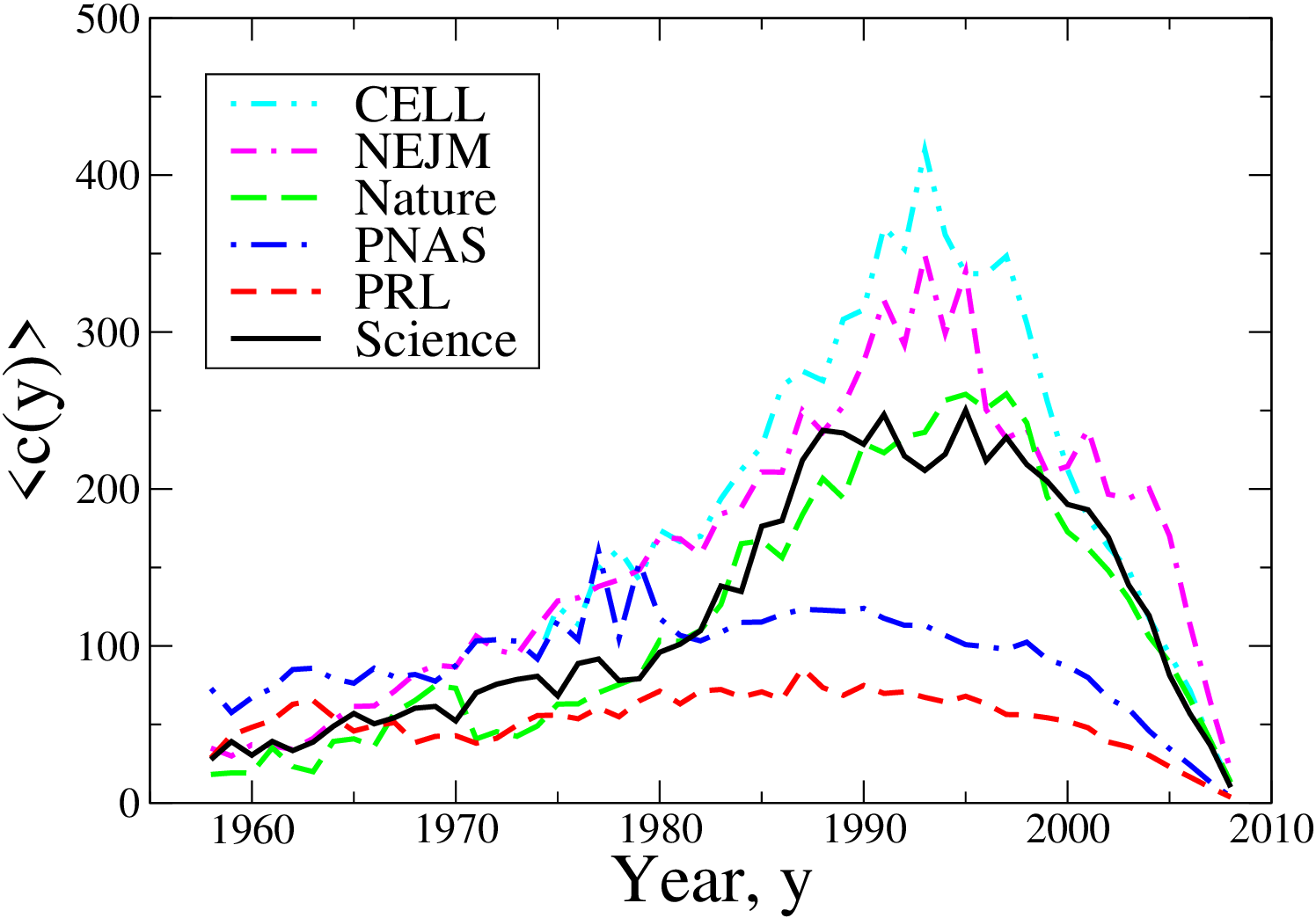}}
%\centering{\includegraphics[width=0.32\textwidth, height=0.25\textwidth]{Journal_y_Csd.eps}}
  \caption{ The average number of citations $\langle c(y) \rangle$ per article for each journal  in year $y$
demonstrates the time-dependence of citations.
  This quantity serves as a normalizing factor, so that we can de-trend citation values across different years. 
The popular {\it Impact Factor} (IF) \cite{Jranking,IFactor} of a journal for a particular year
  is  the average number of citations obtained in a given year for articles published over the previous two years.  In
this paper we restrict our analysis to journals with large IF, ensuring that there is considerable competition  for
limited publication space in such journals. 
  \label{JournalAve} 
}
\end{figure}

\begin{figure}
\centering{\includegraphics[width=0.9\textwidth]{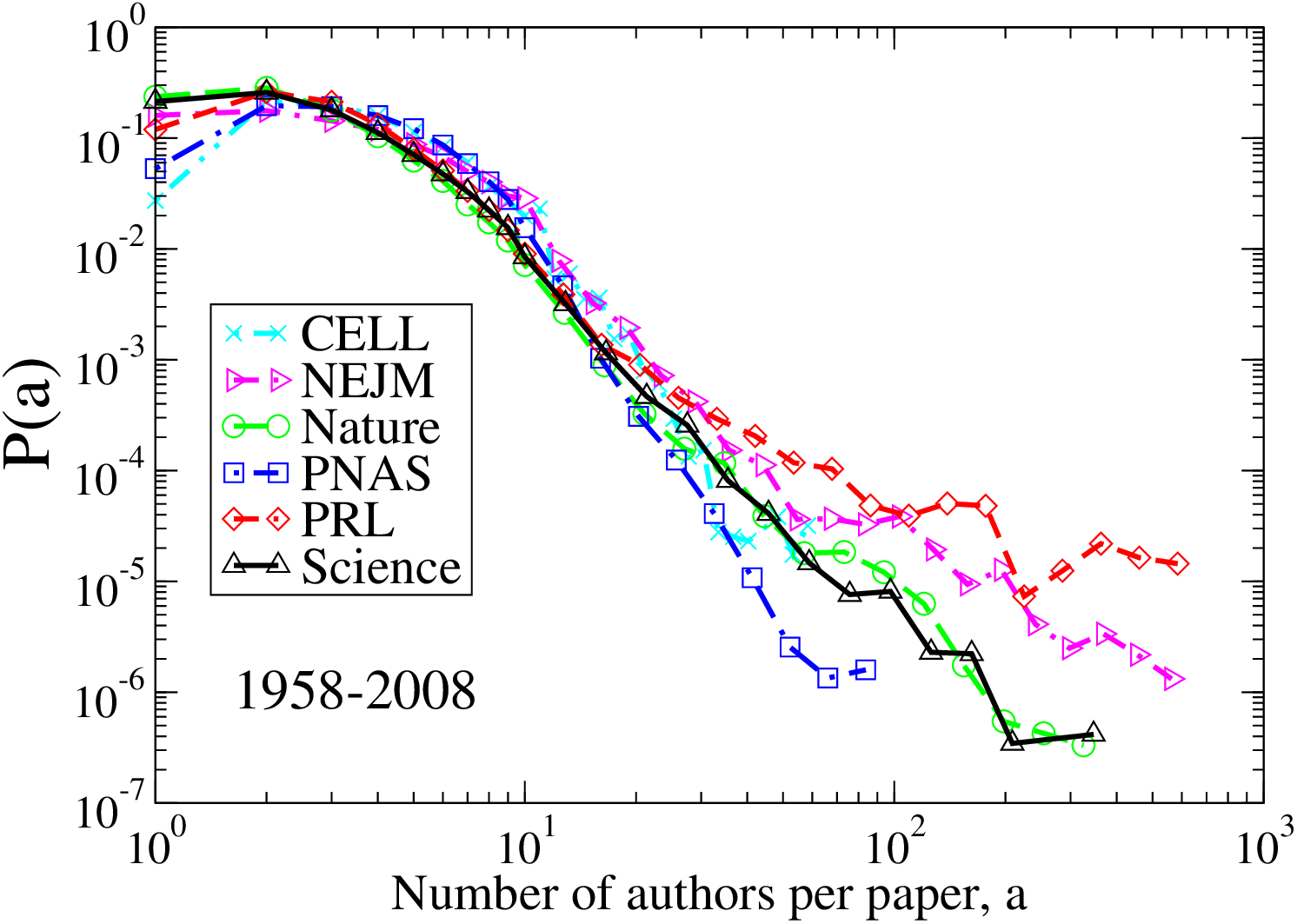}}
  \caption{  The citations and credit for a publication are typically shared fully among all $a$ coauthors, unless the
journal specifically allows for designation of specific credit. 
  The pdf of $a$ for each journal demonstrates that the credit for a single publication can be distributed across a very
broad number of contributors. 
  In this paper, we propose normalizing credit into fractional ``shares", to account for the variations in collaboration
size.
   The average number of authors contributing to an article in each journal are 
  $ \langle a \rangle  =  4.8$ ({\it CELL}), $5.9$ ({\it NEJM}), $3.5$ ({\it Nature}), $4.6$ ({\it PNAS}), $9.1$ ({\it
PRL}), $3.7$ ({\it Science}). 
  \label{nAuthorsPDF} 
}
\end{figure}

Naturally, some older papers will  have more citations than younger papers only because the older papers have 
been in circulation for a longer time. 
In Fig.~\ref{JournalAve} we plot $\langle c(y) \rangle$, the average number of citations for articles from
a given year, and confirm that the time-dependence of citation accumulation is an important factor. 
Interestingly, it is found in  \cite{Jranking} that the pdf of citations from papers within a given year and journal is
approximately log-normal, where the average value of the distribution has a time-dependent drift. With increasing time, 
the pdf approaches a steady state distribution which is also approximately log-normal. Hence, the non-monotonicity in
$\langle c(y) \rangle$ suggests that an important factor in the dynamics of citation counts is the growth with time of
the scientific body and the scientific output.  The mechanism  underlying the evolution of citation trends
and impact factors is complex, where it is found that  citation growth rates decompose into several components in addition to the growth of science \cite{IFactor}. 
Another criticism of Eq.~(\ref{totalcitations}) is that it does not  take into account the variability in number of
coauthors, which varies both within and across discipline (see Figure \ref{nAuthorsPDF}). 

To remedy these problems, we propose a simple success metric termed {\it citation shares}, which normalizes the
citations $c_{i}(y)$ of paper $i$ by $\langle c(y)\rangle$, the average number of citations for papers in a given journal in year $y$,
and divides the quantity $c_{i}(y)/\langle c(y)\rangle$ into equally distributed shares among the  $a_{i}$ 
coauthors. Dividing the shares equally will obviously discount the value of the efforts made by greater
contributors while raising the value of the efforts made by lesser contributors. Without more accurate reporting schemes
on the extent of each authors' contributions (as is now implemented in e.g.  Nature and PNAS), dividing the
shares equally is the most reasonable method given the available data. Hence, we calculate the normalized career
citation  shares as
\begin{equation}
C_{s}=\sum_{i=1}^{n} \frac{1}{a_{i}}  \frac{c_{i}(y)}{\langle c(y)\rangle}\ .
\label{Mciteshares}
\end{equation}
An analogous  estimator for career productivity is $P_{s}$, the total  number of paper shares  within
a given journal, 
\begin{equation}
P_{s}=\sum_{i=1}^{n}\frac{1}{ a_{i}} \ ,
\label{Mpapershares}
\end{equation}
which partitions the credit for each publication
into equal shares among the $a_{i}$ coauthors. 

%We also estimate author career length $L_{z,x}$ within a high-impact journal as the duration 
%in years between an author's last and first article. This definition has limitations for short (career longevity
%$\approx 1$ year) and long (career longevity $\approx $ length of data set $\Delta L= 50$ years). 

There is another sampling bias that we  address. Currently, we assume that all careers are comparable in their duration,
or more precisely, maturity. However, without further consideration, this assumption would ensure that we are comparing
the careers of graduate students with seasoned professors. Hence, we implement a 
 standard method to isolate ``completed"  careers from our data set which begins at year $Y_{0}$ 
and ends at year $Y_{f}$, a common method described in \cite{huber98}. For each author $z$ we calculate $\langle \Delta
\tau_{z} \rangle$, his/her average time  between successive publications in a particular journal. A career which begins
with the first recorded publication in year $y_{z,0}$ and ends with the final recorded publication in year $y_{z,f}$ 
is considered ``complete"  if the following two criteria are met: \\

 \noindent $(i)  \ \ y_{z,f} \leq Y_{f} - \langle \Delta \tau_{z} \rangle$  \ \  and \\
 \noindent $(ii)  \ \ y_{z,0} \geq Y_{0} +\langle \Delta \tau_{z} \rangle$.\\

 \noindent 
 In other words,  this method estimates that the career begins in year $y_{z,0}-\langle \Delta \tau_{z} \rangle$ and
ends in year  $y_{z,f}+\langle \Delta \tau_{z} \rangle$. 
 If either the beginning or ending year do not lie within the range of the data base, then we discount the career as
incomplete to first approximation. Statistically, this 
 means that there is a significant probability that this author published before $Y_{0}$ or will publish after $Y_{f}$. 
Using this criterion reduces the size of the data set by approximately $25\%$ (compare the raw data set sizes $N$ in
Table
\ref{table:journals} to the 
 data set sizes $N^{*}$ in Table \ref{table:CiteSharesCompleteCareers}). The results reported in this paper are, unless
otherwise stated, based on the analysis of only the subset of  ``completed" careers. Another justification for this
criterion is that we recently implement this method in previous work on career longevity in \cite{BB2}, which is more
sensitive to using or not using the criterion.

We note some potential sources of systematic error in the use of this database:
\begin{enumerate}
\item  Degenerate names leads to misleading increases in career totals.
\item  Authors using middle initials in some but not all instances of publication  decreases career totals.
\item  A mid-career change of (last) name   decreases career totals.
\item  Sampling bias due to finite time period. Recent young careers are biased toward short careers. 
Long careers located toward the beginning $Y_{i}$ or end $Y_{f}$ of the database are biased toward short careers and
hence decrease career totals.
\end{enumerate} 
Radicchi {\it et al.} \cite{DiffusionRanking}  observe that the method of concatenated author ID leads to a pdf $P(d)$
of degeneracy $d$ that scales as $P(d) \sim d^{-3}$, which contributes to the systematic error mentioned  in item $1$.
Although the size of our data set guarantees almost surely that such errors exist (given the prevalence of last names
Wang, Lee, Johnson, etc.), these errors should be negligible in the estimation of pdf parameters quantifying a significant portion of the 
data set.

\section{Results}
\label{section:Results}
\subsection{Individual Papers} The growth dynamics of citations vary, ranging from stunted growth to steady growth and,
in some cases where research is published ahead of its time, to late blooming growth \cite{110PhsRev}. One  objective of
this paper is to account for the time-dependence of citation counts in a consistent way so that citations can be
compared across time. We de-trend citation counts to a time-independent framework by dividing the number of citations a
paper has received by the average number of citations for all papers published in the given journal in the same year. 
In Fig. \ref{JournalAve} we plot $\langle c(y)
\rangle$, the average number of citations per paper, where the average is performed over  the full set of papers in 
each given journal  for each year $y$. We note that $\langle c(y)\rangle$ approaches zero as the year becomes
contemporaneous with the data download date, and that
 the peak value of $\langle c(y)\rangle$  occurs for papers  published approximately $15-20$ years before our data
download date in 2009. The presence of this maximum value reflects the growth of the scientific body, the growth of
scientific productivity, and the time delay over which ideas become relevant and established.   See Ref. \cite{Jranking} for the average citation profiles $\langle c(y)
\rangle$ of 2,266 journals indexed by {\it ISI}. Normalizing
to this standard baseline allows one to compare the success of papers across scientific  disciplines, first demonstrated in
 Ref~\cite{citationsPNAS}.

\begin{figure}
\centering{\includegraphics[width=0.75\textwidth]{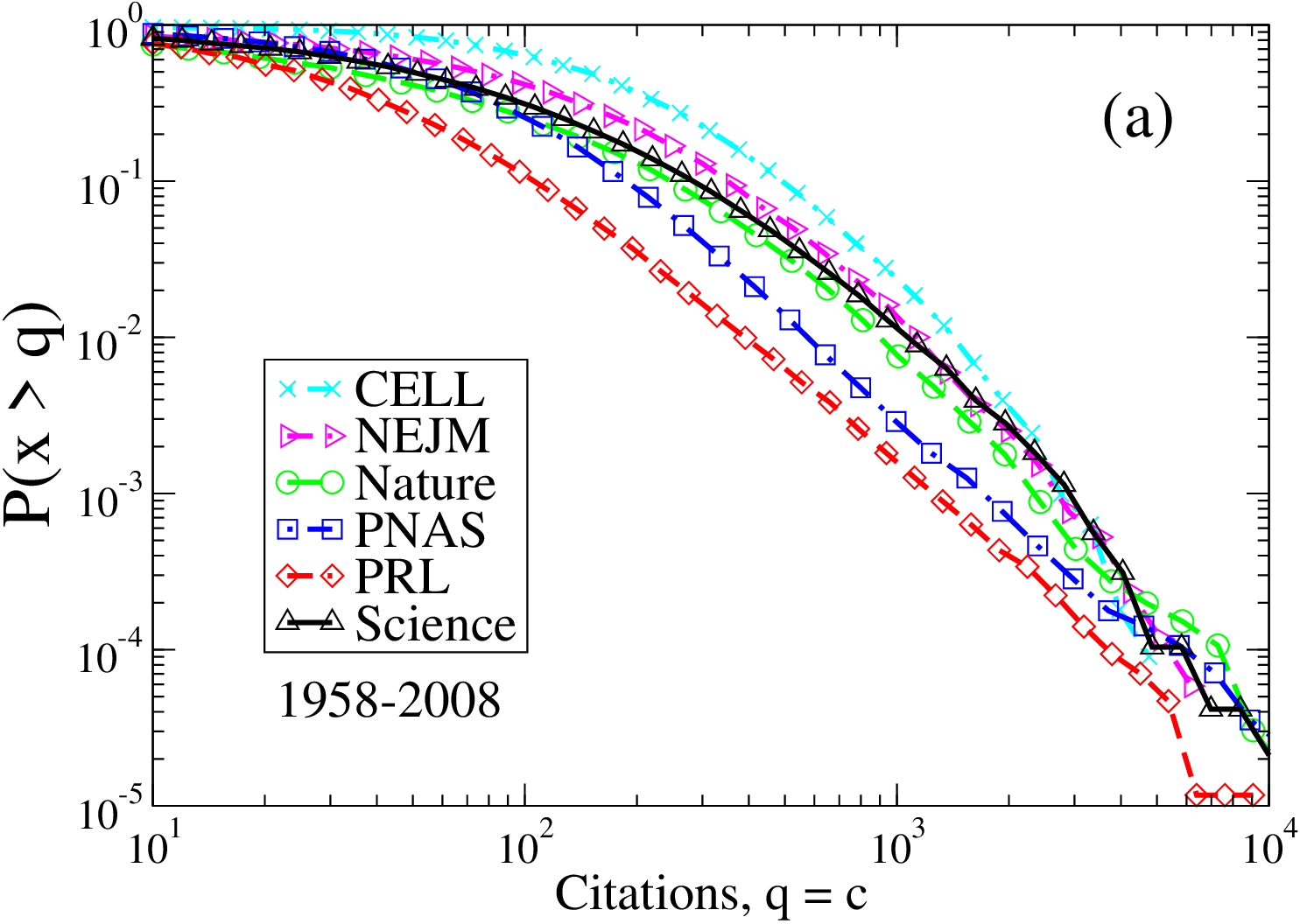}}\\

\centering{\includegraphics[width=0.75\textwidth]{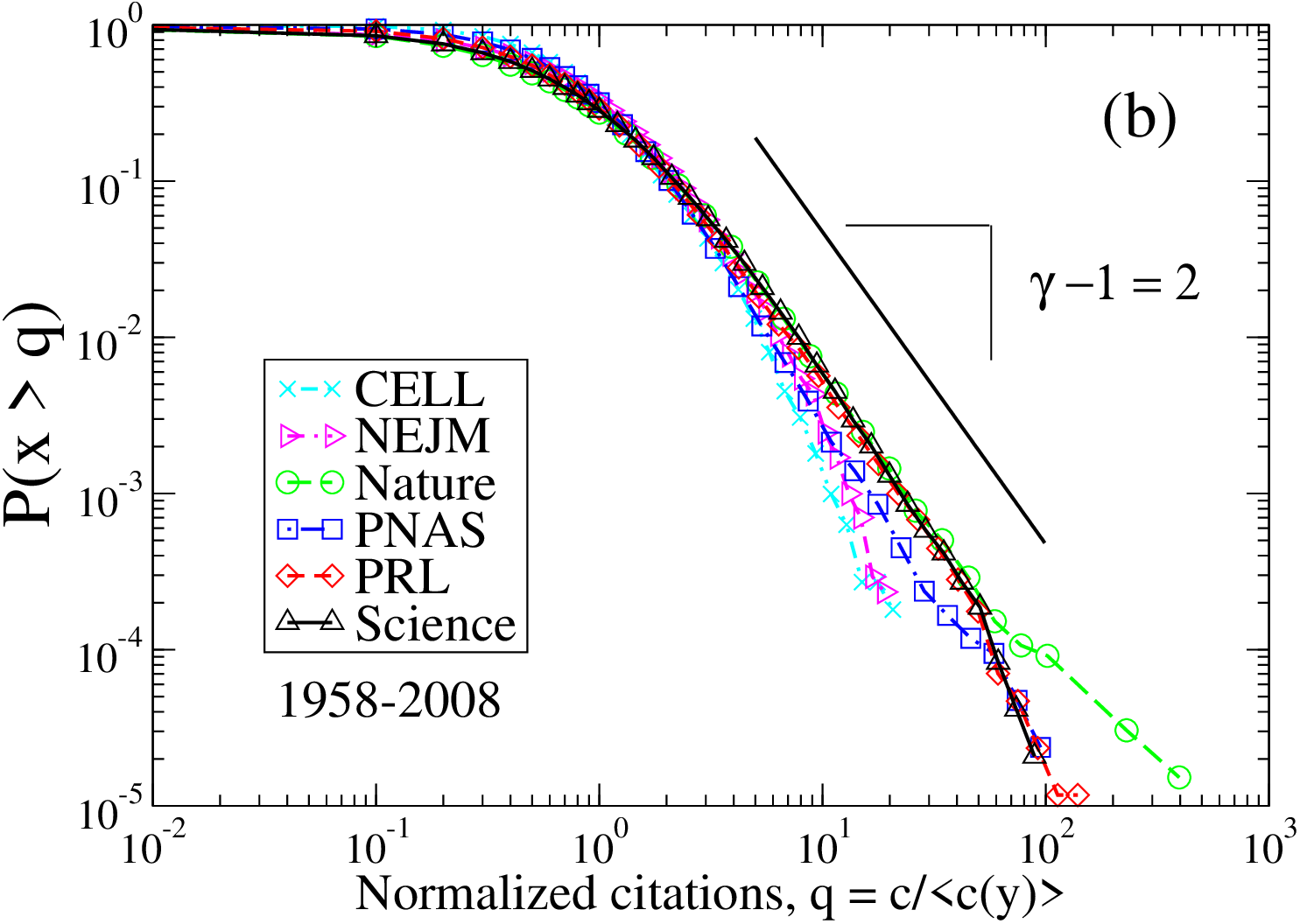}}
  \caption{  Data collapse in the distribution of  article citations  for several journals is achieved by accounting for
the time dependence of 
  citations. (a) The CDF of raw citations $c$ depends on the field. (b)
   Normalizing the number of citations $c$ by $\langle c(y) \rangle$, the average number of citations for a particular
year in a particular journal, the
    CDF for different journals are remarkably similar, with the power-law with $\gamma -1 \approx 2$. We calculate
$\gamma$ for each journal using Hill's MLE and obtain values for the
    scaling exponent corresponding to each journal: $\gamma = 3.64 \pm 0.12$ ({\it CELL}), $3.31 \pm 0.07$ ({\it NEJM}),
$2.87 \pm 0.03$ ({\it Nature}), $3.30 \pm 0.04$ ({\it PNAS}), $2.96\pm 0.03$ ({\it PRL}), $2.86 \pm 0.03$ ({\it
Science}).
    We provide a  power law (solid line) with exponent $\gamma-1=2$ for reference.
  \label{JounalCDF} 
}
\end{figure}

 In order to visualize the effects of normalizing citations to a local average in the case of single papers, we compare
the un-normalized cumulative distribution functions (cdf)  of Fig.
\ref{JounalCDF}(A) with the normalized cdf of Fig. \ref{JounalCDF}(B). The procedure of normalizing by the local average 
reduces the variations across
journal (discipline \cite{DiffusionRanking}), revealing a universal scaling law $P(x>q) \sim q^{1-\gamma}$ with $\gamma -1 \approx 2$. The
scaling exponent $\gamma \approx 3$ describing the 
success of individual papers was first reported in \cite{howpopular}, where normalizing techniques were not employed. 
Surprisingly, we observe the same value of $\gamma$ here for the normalized citation statistics of individual papers
from several major journals over a 50-year period. 

In addition to time-dependent factors, we also consider factors resulting from various research styles across the broad
range of scientific disciplines. In science, the resources required to make significant scientific advances  range from
a pencil and paper to million-dollar laboratory equipment.  Similarly, the number of contributors to scientific advances
ranges from a single scientist to projects involving several hundred scientists. Fig.~\ref{nAuthorsPDF} illustrates the
pdf  of collaboration size associated with a single publication.  The pdf is significantly right-skewed, especially for
publications in NEJM and PRL where occasionally, the number of authors contributing to a single publication
exceeds 100 individuals. For instance, in the cases of research at major medical institutes and particle accelerators,
it is common for the credit for the scientific advance reported in a publication to be shared by extremely large numbers
of contributors. In Eqs.~(\ref{Mciteshares}) and (\ref{Mpapershares}) we choose a simple weighting recipe for
associating credit among $a_{i}$ authors of a single paper $i$.  We  assign equal credit for all authors. Although this
recipe may grant some authors  more credit than due, it also credits other authors with less credit than due. We believe
this weighting scheme is useful in proportionally sharing the credit for a scientific advancement among the $a_{i}$
authors. To address this issue, the  journals  Nature and PNAS require the corresponding author to assign
credit to each co-author across a broad range of categories such as theoretical analysis, experimental methods, and
writing of the manuscript. If adopted across all journals, this formalism could potentially  improve the quantitative
allocation of scientific credit, thus improving the quantitative measures for individual scientific impact.

\subsection{Citation Shares} 
\begin{figure}
\centering{\includegraphics[width=0.75\textwidth]{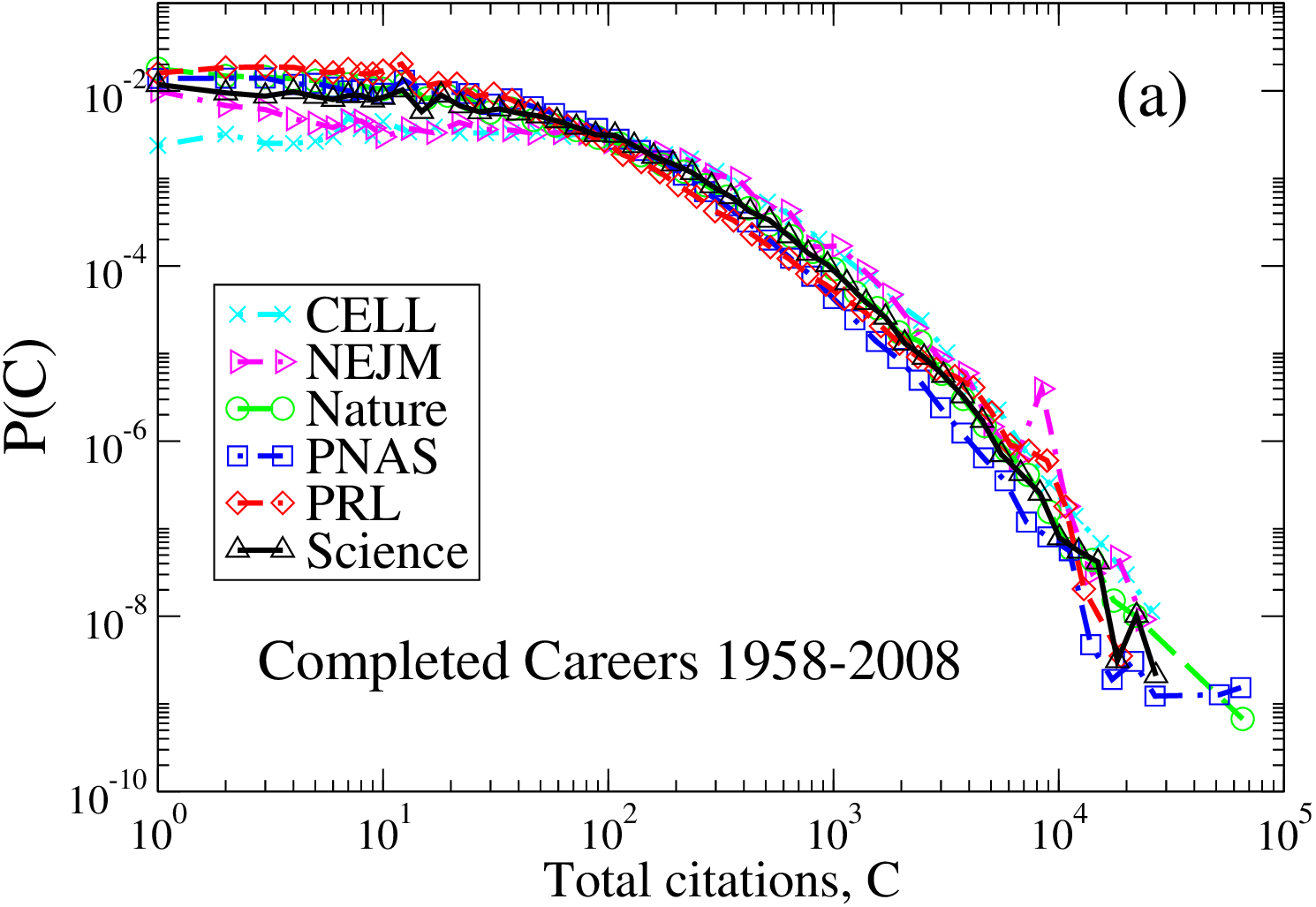}}\\

\centering{\includegraphics[width=0.75\textwidth]{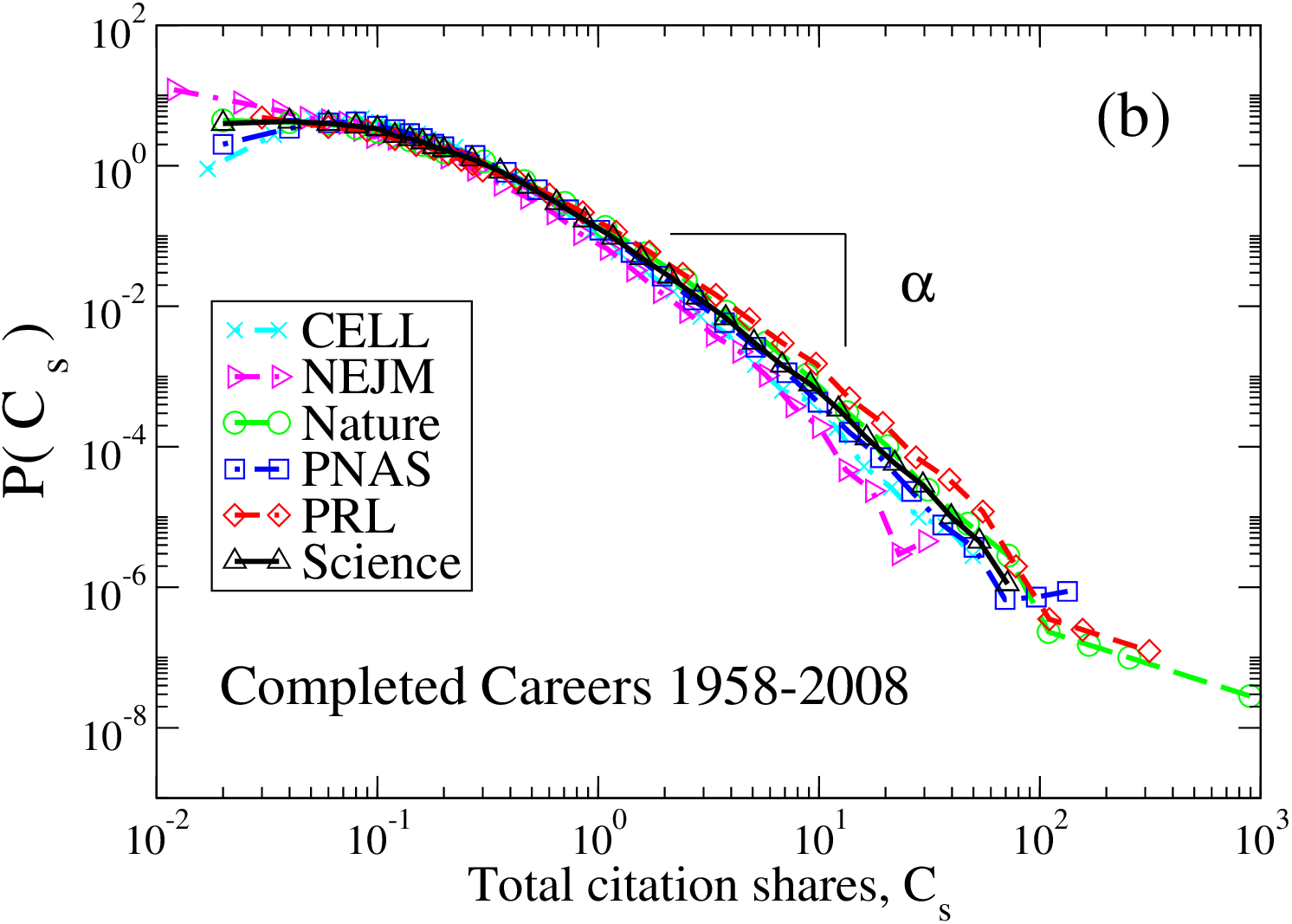}}
  \caption{ 
 We estimate the career success of a scientist within a given journal using the citation shares metric $C_{s}$ defined
in Eq. (2), which accounts for both the number of authors and the age of the paper. 
 (a) PDF of total raw citations $C$ according to Eq.~(\ref{totalcitations}) for ``completed"  careers. (b) PDF of total
citation shares $C_{s}$ according to Eq.~(\ref{Mciteshares}) for ``completed" careers. A given career is considered
``complete"
if there is a large likelihood that the data set  contains all of the particular author's publications.  The
normalization procedure results in significant data collapse in panel (b), with the value of 
the scaling exponent $\alpha \approx 2.5$ for all journals analyzed.
  \label{CiteSharesCDF} }
\end{figure}

In Figs.~\ref{CiteSharesCDF}(A) and ~\ref{CiteSharesCDF}(B) we present the pdfs of career
citations $C$ and of career citation shares $C_{s}$, corresponding to Eq.~(\ref{totalcitations}) and
Eq.~(\ref{Mciteshares}),
respectively. While the six pdfs of  $C$ in Fig.~\ref{CiteSharesCDF}(A) are all similarly
right-skewed, the  collapse onto a universal 
function is weak for small values of $C$. 

The discrepancy between the pdf curves 
for small and large citation counts in Fig.~\ref{CiteSharesCDF}(A) is likely associated with factors associated with the
size of the scientific field, the size of the collaboration group and the impact of the research. Since we study only
six high-impact journals, these factors should be negligible in the overall difference across discipline and journal,
since we assume both discipline and journal are  large. Hence, the collapse of the six pdfs for normalized citation
shares in Fig.~\ref{CiteSharesCDF}(B) demonstrates that normalization is  necessary. For each  pdf $P(C_{s})$  we
observe a scaling regime
\begin{equation}
P(C_{s})\sim C_{s}^{-\alpha} \ ,
\end{equation}
and we estimate the scaling exponent $\alpha$ in the tail of the pdf using the maximum likelihood estimator (MLE), also
known as the Hill estimator \cite{MEJN, ClausetPowLaw}. 

\begin{table}
\caption{ Summary of citation shares  for ``completed" careers. The reduced size of the data set has size $N^{*}$. The
average number of citation shares 
$\langle C_{s} \rangle$ for careers within each journal are computed from the subset of  ``completed" careers (the value
in parenthesis corresponds to value for 
all careers). The value of the  power-law exponent $\beta$ corresponds to the Zipf plot of citation shares plotted in
Fig. \ref{Zipf}, where we calculate the value of $\beta$ using data in the range $10 < rank < N_{MLE}$ implementing a
linear regression  on a log-log scale, where $N_{MLE}$ is the number of data values used to calculate $\alpha$. The
value of the  power-law exponent $\alpha$ corresponds to the pdf  of citation shares plotted in Fig.
\ref{CiteSharesCDF}(B), where we calculate the value of $\alpha$ using Hill's maximum likelihood estimator for data
values greater than a cutoff $C_{s}^{c} \equiv  1$. 
%The expected relationship is $\alpha \approx 1+ 1/\beta$.
}
\begin{tabular}{@{\vrule height 10.5pt depth4pt  width0pt}lc||c||c||c||}\\
%&\multicolumn4c{Least-square values}&\multicolumn3c{gamma distribution moments}\\
\noalign{
\vskip-11pt}
\vrule depth 6pt width 0pt \textbf{\em Journal}  &  $N^{*}$ & $\langle C_{s} \rangle$ & $\beta$ & $\alpha$ \\
\hline 
CELL & 23,060 & \ \  0.34 \ (0.35) \ \ &  \ \ $ 0.52 \pm 0.01$ \ \  & \ \  $2.60 \pm 0.04$ \ \   \\
NEJM &  49,341 & 0.25 \ (0.26)  & $ 0.45 \pm 0.01$ & $2.65 \pm 0.04$\\
Nature & 94,221 & 0.46 \ (0.50) &  $ 0.56 \pm 0.01$ &  $2.42 \pm 0.02$ \\
PNAS &  118,757 & 0.42 \ (0.46) &  $ 0.57 \pm 0.01$ &  $2.50 \pm 0.02$\\
PRL &  72,102 & 0.61 \ (0.75) & $ 0.55 \pm 0.01$ & $2.25 \pm 0.02$ \\
Science & 82,181 &  0.43 \ (0.44)  & $ 0.56 \pm 0.01$ & $2.43 \pm 0.02$ \\
\hline
\end{tabular}
\label{table:CiteSharesCompleteCareers}
\end{table}

We list $\alpha$ values  calculated for $C_{s}> C_{s}^{c} \equiv 1$ in Table~\ref{table:CiteSharesCompleteCareers}. The
Hill estimator is a
robust method for approximating power-law exponents which incorporates each data observation $C_{s}^{i}$ greater than
a cutoff value $C_{s}^{c}$ into the calculation of $\alpha$,
\begin{equation}
\alpha= 1+ \frac{N}{\sum_{i}\ln(C_{s}^{i}/C_{s}^{c})} \ ,
\end{equation}
with standard error,
\begin{equation}
\delta\alpha \approx (\alpha-1)/\sqrt{N} \ .
\end{equation}
For each journal, the the number of data points $N$ greater than $C_{s}^{c}$ used in the calculation of $\alpha$ is approximately
 $10\%$ of the total data set size $N^{*}$. Remarkably, the scaling exponent for citation statistics of completed
careers is approximately $2.5$ for all journals analyzed. Hence, we find convincing evidence for a universal scaling
function representing the distribution of citation shares for scientific careers in competitive high-impact journals.
Interestingly, the values of $\alpha$ for each journal are less than the values of $\gamma \approx 3$ which describes
the scaling of normalized single article citation counts in Fig.~\ref{JounalCDF}. This result implies that the success
of individuals over their entire careers is not related in a simple way to the success of a random number of independent
articles. Instead, there is a larger number of stellar careers than would be expected from the number of stellar papers.

\begin{table}
\caption{ The top 20  authors (not necessarily ``completed'') in the journals  CELL,
NEJM, and  PRL, ranked according to citation shares $C_{s}$ accumulated from their $n$ papers published in each journal. 
Our normalization procedure 
 offers one  way  to quantitatively order the top authors.
 }
\centering{ {\tiny
\begin{tabular}{@{\vrule height .5 pt depth4pt  width0pt}|ccc|ccc|ccc|}
%\begin{tabular}{@{\vrule height .5 pt  << Controls Vertical Spacing between lines !!
\multicolumn3c{{\bf  CELL}}&\multicolumn3c{{\bf  NEJM}} & \multicolumn3c{{\bf  PRL}}\\
\noalign{ 
\vskip-1pt} Name & $C_{s}$ & $n$  & Name & $C_{s}$ & $n$ & Name & $C_{s}$ &
$n$  \\
\hline 
GREEN, H & 49.7 & \ \ \ 35 \ \ \ & BRAUNWALD, E & 30.3 & \ \ \ 59 \ \ \ & WEINBERG, S & 313.3 & \ \ \ 49 \ \ \   \\
BALTIMORE, D & 33.8 & 64 & KOCHWESER, J & 23.6 & 28 & ANDERSON, PW & 137.4 & 64 \\
MANIATIS, T & 29.5 & 55 &  MCCORD, JM & 20.2 & 1 & WILCZEK, F & 120.0 & 62 \\
SHARP, PA & 25.1 & 41 & FINLAND, M & 17.4 & 36 & TERSOFF, J & 105.1 & 76 \\
TJIAN, R & 23.8 & 45 & HENNEKENS, CH & 16.9 & 36 & HALDANE, FDM & 102.3 & 38 \\
LEDER, P & 22.4 & 39 & REICHLIN, S & 16.7 & 10 & YABLONOVITCH, E & 87.5 & 21 \\
AXEL, R & 20.9 & 52 & VECCHIO, TJ & 14.8 & 1 & PERDEW, JP & 78.3 & 20 \\
WEINTRAUB, H & 20.5 & 46 & STAMPFER, MJ & 14.3 & 45 & LEE, PA & 74.6 & 76 \\
KARIN, M & 18.5 & 40 & TERASAKI, PI & 13.7 & 29 & PENDRY, JB & 74.1 & 29 \\
RUBIN, GM & 18.0 & 52 & OSSERMAN, EF & 13.7 & 6 & PARRINELLO, M & 72.8 & 68 \\
KOZAK, M & 17.1 & 6 & KUNIN, CM & 13.5 & 16 & FISHER, ME & 71.6 & 67 \\
ROEDER, RG" & 15.5 & 44 & YUSUF, S & 13.4 & 18 & CIRAC, JI & 66.7 & 97 \\
RHEINWALD, JG & 14.7 & 7 & ROSEN, FS & 13.2 & 42 & HALPERIN, BI & 66.7 & 50 \\
EVANS, RM & 14.1 & 32 & CHALMERS, TC & 13.1 & 30 & RANDALL, L & 63.4 & 14 \\
OFARRELL, PH & 13.9 & 14 & AUSTEN, KF & 12.9 & 30 & BURKE, K & 63.2 & 18 \\
GLUZMAN, Y & 13.3 & 2 & WELLER, TH & 12.7 & 7 & JOHN, S & 62.8 & 20 \\
HUNTER, T & 13.2 & 27 & GARDNER, FH & 12.6 & 19 & GEORGI, H & 61.9 & 26 \\
GOLDSTEIN, JL & 13.0 & 36 & DIAMOND, LK & 12.6 & 18 &  CAR, R & 59.8 & 51 \\
PENMAN, S & 12.9 & 30 & FEINSTEIN, AR & 12.2 & 16 &  GLASHOW, SL & 59.6 & 37 \\
BROWN, MS & 12.8 & 35 & MERRILL, JP & 11.9 & 25 & CEPERLEY, DM & 58.9 & 39 \\
%"BERK" "AJ" & 12.6 & 7 & "WILLETT" "WC" & 11.4 & 43 & "SUNDRUM" "R" & 58.7 & 4 \\
\hline
\end{tabular}}}
\label{table:champions}
\end{table}

\begin{figure}
\centering{\includegraphics[width=0.9\textwidth]{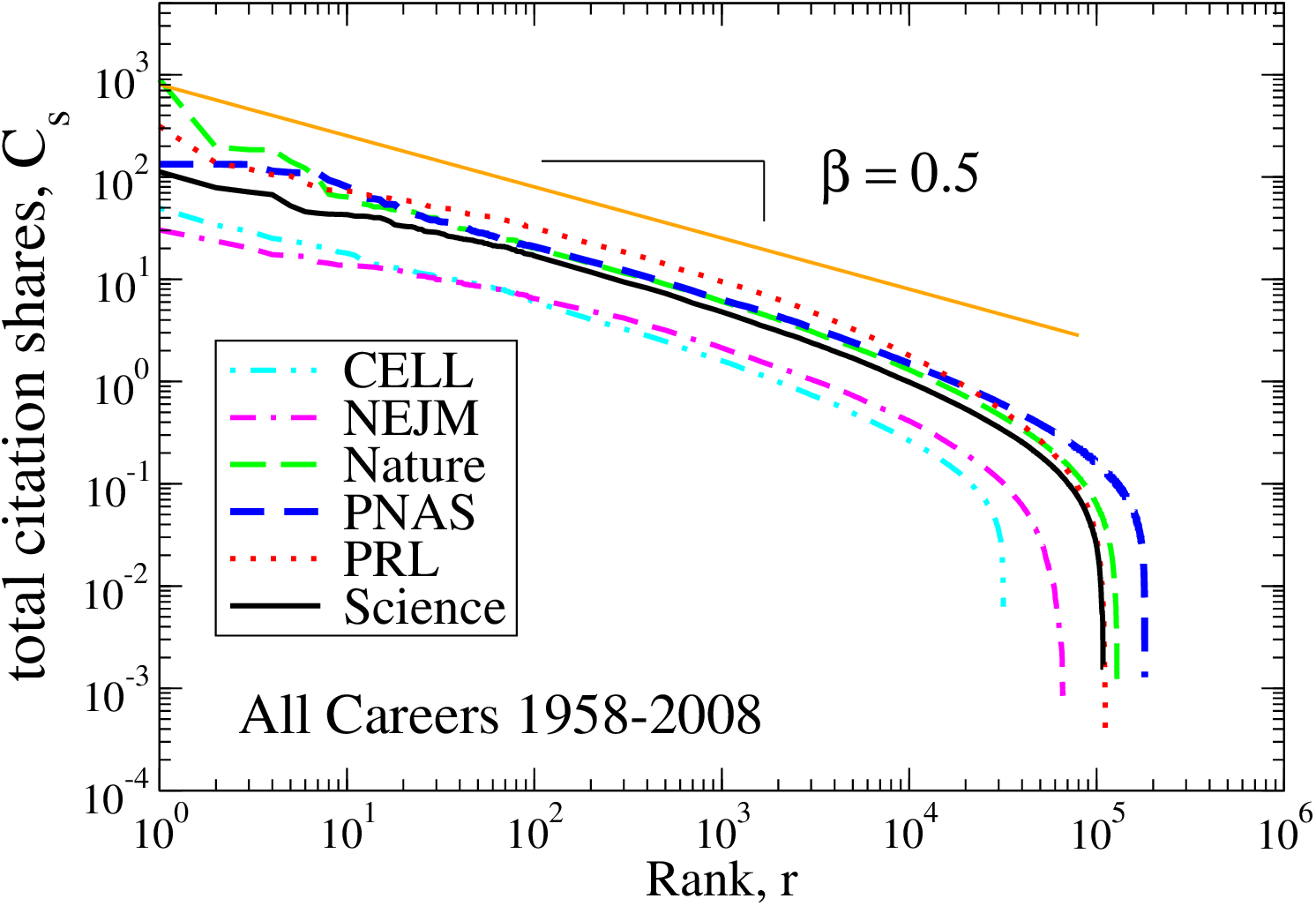}}
\caption{  The Zipf plot emphasizes the stellar careers corresponding to large $C_{s}$, the total number of citation
shares within a particular journal defined in Eq. (2), and shows a significant scaling regime corresponding to the
top-ranking ``champions" of each journal. 
For comparison, we list the top 20 careers within the journals {\it CELL}, {\it NEJM} and {\it PRL} in Table
\ref{table:champions}. 
The total number of career citation shares for a particular author in a given journal serves as a proxy for the career
success of the scientist. The
 statistical regularity in the rank ordering of scientific achievement extends over four orders of magnitude. The
similarity in scaling exponent among the journals analyzed possibly suggests that there 
 are fundamental forces governing success in competitive arenas such as high-impact journals. For visual clarity, we
plot the power law with scaling exponent $\beta \equiv 0.5$. 
\label{Zipf}}
\end{figure}

Another illustrative method for comparing the distribution of success across the entire range of individuals is the
popular Zipf plot, which is mathematically related to the pdf \cite{howpopular,BrazilPhysChem}. In Fig. \ref{Zipf} we
plot $ C_{s}$ versus rank for the same set of  completed careers analyzed in Fig. \ref{CiteSharesCDF}(B). The
Zipf plot emphasizes the scaling in the tail of the pdf, which are represented by high rank values. We calculate
the scaling exponent of the Zipf plot for rank values  in the range $10 < r < r_{c}$ for each journal, where $r_{c}$
corresponds to the
 number of data points incorporated into the calculation of 
$\alpha$ using Hill's MLE. These values are in approximate agreement with the expected relationship $1+1/\beta
\approx \alpha$.

 The small range of  $\beta$ values across journals (see Table~\ref{table:CiteSharesCompleteCareers}) demonstrates
 that our normalization procedure places scientific accomplishments on a comparable footing across both time and
discipline. In Table~\ref{table:champions} we list the top 20 publication careers   according to citation shares. This
table consists mostly of careers that have many papers of significant impact; however, it also contains a  few careers
that are distinguished by a small number of seminal papers. Hence, while longevity at the upper tier of science is good
at assuring reputation and success, there are also a few  instances of success achieved via a singular yet monumental
accomplishment. 

\subsection{Paper Shares} 
\begin{figure}
\centering{\includegraphics[width=0.9\textwidth]{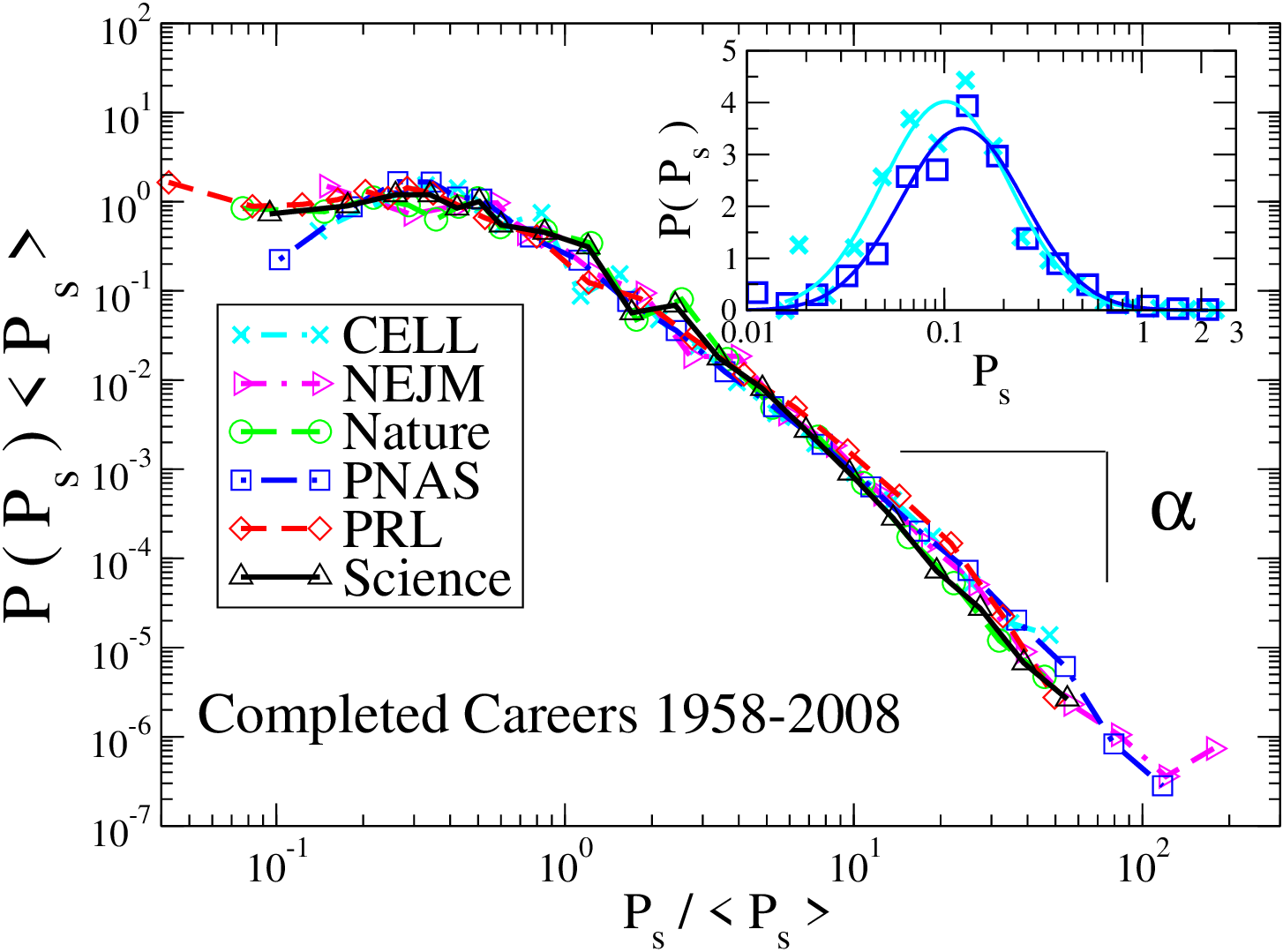}}
%\centering{\includegraphics[width=0.32\textwidth, height=0.25\textwidth]{Journal_y_Csd.eps}}
  \caption{   As a proxy for career productivity, we define paper shares $P_{s}$ in Eq.(\ref{Mpapershares}), which accounts for
variations in the size of the collaboration. 
  In order to collapse the pdfs of total paper shares for completed careers within the six journals analyzed, we
hypothesize that the universal scaling function quantifying productivity can be written as $P(P_{s})=f(P_{s}/\langle P_{s} \rangle)$.
We approximate the generic pdf of paper shares 
 by a log-normal distribution with a  power-law tail after a cutoff value $P^{c}_{s} \approx 1$. We list the values of the log-normal  parameters $\mu$ and $\sigma$, and the scaling parameter $\alpha$ for each journal in Table \ref{table:PaperSharesCompleteCareers}. (Inset) 
 We plot the pdf for {\it CELL} and {\it PNAS} data on log-linear axes for $P_{s} \leq 3$ in order to
demonstrate the log-normal form consistent with the prediction by W. Shockley \cite{ShockleyProductivity} that
productivity can be modelled as a series of random multiplicative factors. 
  \label{PaperSharesCDF} }
\end{figure}

We now focus on scientific productivity, quantified by the number of papers published by a given 
author. In Fig.~\ref{PaperSharesCDF} we plot the  pdfs for  paper shares defined in 
Eq.(\ref{Mpapershares}).  In order to collapse the pdfs for the six journals analyzed, we hypothesize that a universal
function for productivity can be written as $P(P_{s})=f(P_{s}/\langle P_{s} \rangle)$. 
In an effort to compare the pdfs across discipline, we approximate the generic pdf of paper shares
 by a log-normal distribution with a heavy tail after a cutoff value $P^{c}_{s}$. 
 Quantitatively, we represent the general form of the pdf as,
\begin{equation}
\label{papersharespdf}
 P(P_{s}) \propto \left\{
\begin{array}{cl}
    \frac{1}{P_{s}} \exp[-(\ln P_{s}- \mu)^{2}/ 2 \sigma^{2}] \ \ \ & P_{s}  \lesssim P^{c}_{s} \\
     P_{s}^{-\alpha}\  \ \ &  P_{s} \gtrsim  P^{c}_{s}  \\
 %     \frac{1}{x \sqrt{2\pi \sigma^{2}}} exp[-(\ln x- \mu)^{2}/ 2 \sigma^{2}] \ \ \ & x  \lesssim x_{c} \\
   % \frac{\delta}{p_{c}\Gamma(1/\delta)}  exp[-(x/p_{c})^{\delta}]\  \ \ &  x \gtrsim x_{c} \ . \\
           \end{array}\right.
\end{equation}
The least-square  parameters for the log-normal fit and the MLE parameter for the scaling regime are listed in Table
\ref{table:PaperSharesCompleteCareers}. The log-normal
distribution is consistent with the prediction by Shockley \cite{ShockleyProductivity} that productivity (as estimated
here by paper shares) is a result of a series of multiplicative factors, which can lead to log-normal
\cite{MultiplicativeI}, stretched exponential
\cite{MultiplicativeII}, and even power-law  \cite{ReedH} distributions, given the appropriate set of systematic
conditions.

\begin{table}
\caption{  Summary of paper shares  for ``completed" careers . 
 The value of the  log-normal fit parameters $\mu$ and $\sigma$ correspond to the pdf  before the cutoff value of
$P^{c}_{s} \approx 2$  paper shares. The values of $\alpha$ are calculated using a data values after the cutoff
$P^{c}_{s} \equiv 1$ paper shares, which corresponds to approximately $8\%$ of the total data for each journal. 
}
\begin{tabular}{@{\vrule height 10.5pt depth4pt  width0pt}lc||c||c||}\\
%&\multicolumn4c{Least-square values}&\multicolumn3c{gamma distribution moments}\\
\noalign{
\vskip-11pt}
\vrule depth 6pt width 0pt \textbf{\em Journal}  & $  \mu$  & $\sigma$ & $\alpha$ \\
\hline 
CELL & \ \ $-1.7 \pm 0.1$ \ \ & \ \ $0.7 \pm 0.1$ \ \ & \ \ $2.60 \pm 0.05$ \ \  \\
NEJM &  $-1.7 \pm 0.1$ & $1.0 \pm 0.1$  & $2.60 \pm 0.02$   \\
Nature & $-1.3 \pm 0.1$ & $1.0 \pm 0.1$ &  $2.74 \pm 0.05 $ \\
PNAS &  $-1.6 \pm 0.1$ & $0.7 \pm 0.1$ & $2.56 \pm 0.02$\\
PRL &  $-1.1 \pm 0.1$ & $1.0 \pm 0.1$ & $2.35 \pm 0.02$ \\
Science &$ -1.4 \pm 0.1$ & $0.9  \pm 0.1$ &  $2.61 \pm 0.02$ \\
\hline
\end{tabular}
\label{table:PaperSharesCompleteCareers}
\end{table}

\subsection{Matthew Effect} 
\begin{figure}
\centering{\includegraphics[width=0.75\textwidth]{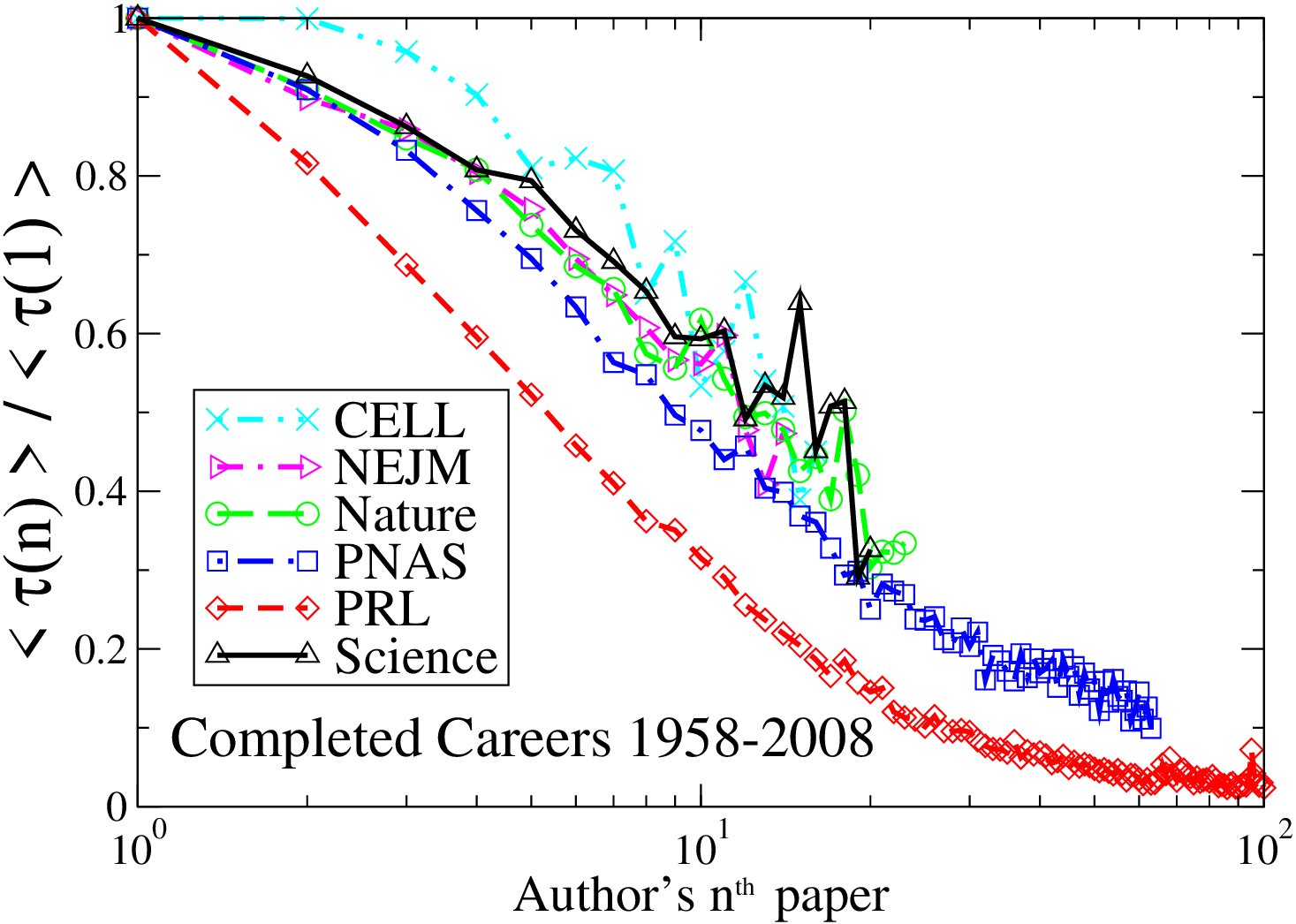}}
%\centering{\includegraphics[width=0.75\textwidth]{Fig7b.eps}}
  \caption{ 
 A decreasing waiting time $\tau(n)$ between publications in a given journal suggests that a longer publication career
(larger $n$) facilitates future publications, as predicted by the Matthew effect.  We plot $\langle
\tau(n) \rangle / \langle \tau(1) \rangle$, the average waiting time $\langle \tau(n) \rangle$ between 
paper $n$ and paper $n+1$, 
rescaled
 by the average waiting time between the first and second publication, $\langle \tau(1) \rangle$ . 
The values of $\langle \tau(1) \rangle$  are $2.2$ ({\it CELL}, {\it PRL}), $3.0$ ({\it Nature}, {\it  PNAS}, {\it
Science})  and $3.5$ ({\it NEJM}) years. {\it Physical Review Letters} exhibits a more rapid decline in $\tau(n)$,
reflecting the rapidity of successive publications (often by large high-energy experiment collaborations), which is
possible in this high-impact letters journal.
 \label{Tx} 
}
\end{figure}

We conclude this section with quantitative evidence of the ``Matthew effect" in the
advancement of scientific careers. 
In Fig. \ref{Tx} we plot  the average waiting time between publications $\langle \tau(n) \rangle$ for all authors that meet
the complete career criterion by averaging the difference in publication year for the  paper $n$ and the paper $n+1$.
The values of $\langle \tau(1) \rangle$ for each journal are $2.2$ (CELL, PRL), $3.0$ (Nature, PNAS, Science)  and $3.5$
(NEJM) years.  The decrease in waiting time between publications is a
signature of the cumulative advantage mechanism qualitatively described in \cite {huberCum} and quantitatively analyzed in
\cite{cumadvprocess, BB2}. 
To avoid presenting statistical fluctuations arising from the small size of data sets,  we only present 
$\langle
\tau(n) \rangle$ computed for data sets exceeding $75$ observations. 

To explain the steady decline of the curve for PRL we mention that PRL has many authors with many articles ($n \gtrsim
100$). A possible explanation is that a significant number of these authors are involved in large particle accelerator
experiments with multiple collaborating groups. These multilateral projects contribute significantly to 
the heavy tail observed in the pdf of the number of authors per paper  (Fig. \ref{nAuthorsPDF}). Hence, the decay in the
curve for PRL which approaches zero might be due to the project leaders at large experimental institutions which produce
over many years many significant results per year. Furthermore, the organization of the curves in Fig. \ref{Tx} suggests
that it is more difficult at the beginning of a career to repeatedly publish in CELL than PRL. 
Reaching a crossover point along the career ladder is a generic phenomenon observed in many professions.
Accordingly, surmounting this abstract crossover is motivated  by significant personal incentives, such as salary
increase, job security, and managerial responsibility.

\section{Discussion}
\label{section:Discussion} 
Scientific careers share many qualities with other competitive careers, such as the careers of professional sports
players, inventors, entertainers, actors, and musicians \cite{BB1, Hollywood, musicians}. Limited resources such as
employment, salary, creativity, equipment, events, data samples, and even individual lifetime contribute to
the formation of 
generic arenas for competition. Hence, of interest here is the distribution of success and productivity in high impact
journals 
which in principle have high standards of excellence. 

In science, there are unwritten guides to success requiring ingenuity, longevity, and publication. 
We observe a quantifiable statistical regularity describing publication careers of individual scientists
across both time and discipline.  Interestingly, we find that the scaling exponent for individual papers ($\gamma \approx 3$) is 
larger than the scaling exponent for total citation shares ($\alpha \approx 2.5$) and the scaling exponent for total paper shares ($\alpha \approx 2.6$), which indicates that there is a higher frequency of 
stellar careers than stellar papers. This is consistent with the observation that a stellar career can result from an arbitrary combination of stellar papers and consistent success, 
as demonstrated in Table \ref{table:champions}. 
In all, the statistical regularity found in the
distributions for both citation shares and paper shares lend naturally to 
methods based on extreme statistics in order to distinguishing stellar careers. Such methods have been developed for 
Hall of Fame candidacy in baseball \cite{BB2, BB3}, where statistical benchmarks are established using the distribution of success. 

Statistical physicists have long been interested in complex interacting systems, and are beginning to succeed in
describing social dynamics using models that were
 developed in the context of concrete physical systems \cite{SocialDynamics}. 
This study is inspired by the long term goal of using quantitative methods from statistical physics to answer
traditional questions rooted in social science \cite{CompSocScience}, such as the nature of competition, success,
productivity, and the universal features of human activity. 
Many studies begin as
 empirical descriptions, such as the studies of common mobility patterns \cite{mobility},
sexuality\cite{sexpowlaw,internetdating}, and financial fluctuations \cite{Mantegna}, and lead to a better understanding
of the underlying mechanics. It is possible that the empirical laws reported here will motivate useful descriptive
theories of success and productivity in competitive environments.

\section*{ACKNOWLEDGMENTS} We thank S. Miyazima and S. Redner for helpful comments and the NSF for financial support.

\newpage
\clearpage

\begin{thebibliography}{99}
\bibitem{authorname} In the ISI database, the author name John M. Doe would be stored as ``Doe, J" or ``Doe, JM"
depending on the author's designation.
\end{thebibliography}
\end{document}